\documentclass[aps,epsf,preprint]{revtex4}
\usepackage{graphicx}% Include figure files
\usepackage{bm}% bold math
\begin{document}
\newcommand{\Sv}{{\bm S}}
\newcommand{\eF}{{\epsilon_F}}
\newcommand{\sz}{{|0\rangle}}
\newcommand{\so}{{|1\rangle}}
\newcommand{\sigmav}{{\bm \sigma}}
\newcommand{\AND}{{\rm AND}}
\newcommand{\XOR}{{\rm XOR}}
\newcommand{\xv}{{\bf x}}
\newcommand{\Mv}{{\bf M}}
\newcommand{\Av}{{\bf A}}
\newcommand{\av}{{\bf a}}
\newcommand{\ra}{{\rightarrow}}
\newcommand{\tr}{{\rm tr}}
\title{
Quantum toys for quantum computing: 
persistent currents controled by spin chirality
} 
\author{Gen Tatara}
\author{N. Garcia}
\affiliation{
Graduate School of Science, Osaka University, Toyonaka Osaka 560-0043, 
Japan}

\affiliation{
Consejo Superior de Investigaciones Cientificas
Serrano 144, 28006 Madrid, Spain
}
%PACS numbers: 73.23.-b, 73.23.Ra, 03.65.Vf
\date{\today}
\begin{abstract}
Quantum devices and computers will need operational units in different
 architectural configurations for their functioning.
The unit should be a simple ``quantum toy'', easy to handle 
 superposition states.
Here a novel such 
unit of quantum mechanical flux state (or persistent current)
in a conducting ring with three ferromagnetic quantum dots is presented.
The state is labeled by the two direction of the persistent current,
 which is driven by the spin chirality of the dots, 
and is controled by the spin. 
It is demosntrated that by use of two rings connected,
one can carry out unitary transformations on
 the input flux state by controling one spin in one of the rings, 
unabling us to prepare superposition states.
The flux is shown to be a quantum XOR operation gate, and may be 
useful in quantum computing. 

\end{abstract}
\maketitle

Realization of quantum mechanical two-level systems and controling the
superposition of the states in experiment is a fundamental but also
an interesting subject.
Such systems are intensively studied recently, since controlling them
is a starting point of the realization of quantum 
computers\cite{Cirac01}.
Such two-level systems, called Qubits, has been implemented, for
instance, in ion 
traps\cite{Turchette95}, nuclear spins\cite{Gershenfeld97},
 and in Josephson junctions\cite{Wal00}.
In the case of flux in Josephson junction, the two-level states are
states with persistent currents in a superconducting loop with
different directions.
The current is induced by a magnetic flux through the ring, and the
quantum superposition of the two current states was observed recently 
by a fine tuning of the flux\cite{Wal00}.

In this paper, we present a novel quantum mechanical flux state, which
is controled by controling the spin in a quantum dot.
The flux here is due to a persistent current in a conducting ring, but
of different origin as Josephson Qbit; namely, current induced by spin 
chirality. 
By putting three (or more) quantum dots which carries quantum spin, 
we show that the wave function of the  flux is 
controled  by that of the spins.
The realization of the superposition state of flux is thus realized
simply by creating a superposition state on one of the spins.
We also demonstrate that this system can be used to 
create entangled states of two or more spins.
This ``quantum toy'' also works as a quantum XOR logic gate, which may
be useful in quantum computers.
We also discuss more sophisticated case of two rings coupled, where we
can carry out untary transformations on the current state.

%%%%%%%%%%%%%%%%%%%%%%%%%%%%%%%%%%%%%%%%

The existence of the spontaneous 
current in a small ring in contact with
three or more ferromagnets when the three magnetization vectors 
form a finite solid angle was pointed out recently in ref. \cite{TK03}. 
The effect is due to the breaking of the 
time-reversal symmetry in the orbital motion
as a consequence of non-commutativity of the spin algebra, and it is
essential that the electron wave function is coherent over the ring.
The current was shown 
to be proportional to the non-coplanarity (spin chirality) 
of the three magnetizations, $(\Sv_1\times\Sv_2)\cdot\Sv_3$,
where magnetizations are represented by classical vectors $\Sv_{1}$, 
$\Sv_{2}$, and $\Sv_{3}$.
%%%%%%%%%%%%%%%%%%%%%%%%
\begin{figure}[bthp]
%\section{Figfnf}
\includegraphics[scale=0.5]{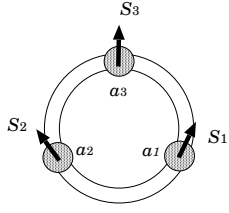}
\caption{
The system of chirality-driven persistent current with three
 ferromagnetic dots.
\label{FIGring}}
\end{figure}
%%%%%%%%%%%%%%%

Here we condider the case where the magnetization is quantum spin of
$S=1/2$, which is 
carried by ferromagnetic dots on the ring, in which case the same reasoning as
in ref. \cite{TK03} applys.
We do not consider the screening due to the Kondo effect, 
considering the temperature higher than the Kondo temperature.
The spins in the dots can then be regarded as qubits.
Note that the decoherence time 
of the electron spin is known to be much larger in general in nanostructures
than that for the charge due to the smallness
of the spin-orbit coupling\cite{Loss97}.
We treat perturbatively the coupling between the conduction electron and 
the spins in the dots.
The equilibrium current at $x$is calculated from 
$J(x)=\frac{e\hbar}{2m}{\rm Im}{\rm Tr} [(\nabla_x -\nabla_{x'}) 
G(x,x',\tau=-0)|_{x=x'}]$, where 
$G(x,x',\tau)\equiv -\langle T c(x,\tau)c^\dagger (x',0) \rangle$ 
is the thermal Green function, and trace is over the spin indices.
The interaction with the spins in the dots can be 
expressed by the potential 
$V(x)=-\Delta \Sv(x) \cdot \sigmav$, where 
$\Sv(x)\equiv \sum_i \hat\Sv_i \delta(x-a_i)$, 
$a_i$ being the position of ferromagnetic dots ($i=1,2,3$), and $\Delta$
represents the effective coupling between the electron and quantum spin,
$\hat\Sv$.
The Green function is determined by the Dyson equation, 
$G=g+gVG$, where the free Green function is denoted by $g$.
By noting that the free Green function is symmetric under spatial
reflection, $g(x,x')=g(x',x)$, and by summing over a path contributing
to the current and its time-reversed path,
the contribution to the current $J(x)$ at $n$-th order in $V$ is shown to be
proportional to
\begin{equation}
\sum_{x_i}{\rm Tr} [V(x_1)V(x_2)\cdots V(x_n)-V(x_n)\cdots V(x_2)V(x_1)] 
\nabla_x g(x,x_1) g(x_1,x_2)g(x_2,x_3) \cdots g(x_n,x).
\end{equation}
The second term in the square bracket corresponds to the contribution
from the time-reversed path.
Since ${\rm Tr}[V(x_1)V(x_2)-V(x_2)V(x_1)]
=\Delta^2 S^\mu(x_1) S^\nu(x_2) {\rm Tr}[\sigma^\mu\sigma^\nu]=0$, 
we immediately see that the leading contribution is 
from the third order with $x_i \in F_i$, which reads 
\begin{equation}
\hat J(x)=
\frac{e\hbar}{m} \int
\frac{d\omega}{2\pi} f(\omega) \nabla_x {\rm Im} [g_{x1}g_{12}g_{23}g_{3x'}] 
|_{x'=x}
4\Delta^3 (\hat\Sv_1\times\hat\Sv_2)\cdot\hat\Sv_3,
\end{equation}
where we have used
${\rm Tr}[\sigma_i\sigma_j\sigma_k] =2i\epsilon_{ijk}$,
$f(\omega)$ is the Fermi distribution function and 
$g_{ij}=g^r(a_i-a_j,\omega)$ ($i,j=x,1,2,3$) is the retarded free Green 
function. 
In the case of one-dimensional ring, the result is
\begin{equation}
  \hat{J} = J_0 \hat{C}_3
\end{equation}
where 
$\hat{C}_3\equiv (\hat{\Sv}_1\times\hat{\Sv}_2) \cdot\hat{\Sv}_3
$ and
 $J_0=-2e \frac{v_F}{L} 
  \cos(k_F L) 
       \left( \frac{\Delta}{\epsilon_F} \right)^3 $.\cite{qspins}
The state of the system is thus specified by a combination of states of the
spin-qubits $\hat{\Sv}_i$ and  a current-qubit $\hat{J}$.
The current takes a  value according to the ''volume''
of the three spins, $(\hat\Sv_1\times\hat\Sv_2) \cdot\hat\Sv_3$.
The magnitude $J_0$ of the present persistent current is different from the
conventional one 
due to a magnetic flux through the ring\cite{Byers61,Buttiker83} 
by a factor of
$(\frac{\Delta}{\epsilon_F})^3$.
The appearance of the current is due to the symmetry breaking of the
charge (U(1)) sector, as in the case of the current in 
Josephson junction.
But note that here the U(1) symmetry breaking was due to the
non-commutativity of spin (SU(2)) sector.

Classically, spin chirality $C_{3}\equiv(\Sv_1\times\Sv_2)\cdot\Sv_3$
(with $\Sv_i$'s as classical vectors) vanishes if any of the $\Sv_i$'s 
are parallel to each other, and is thus 
read as a XOR operation. To be explicite, we choose $\Sv_3//z$, and
then $C_3=\frac{1}{2}(S_1^x S_2^y-S_1^yS_2^x)$.
If we label the state $\Sv_i=\frac{1}{2}\hat x$ as 0 and 
$\Sv_i=\frac{1}{2}\hat y$ as 1, 
the result of
$C_3$ is written as $C_3(00)=C_3(11)=0$, $C_3(01)=-C_3(10)=\frac{1}{8}$
(states are labelled by $(\Sv_1 \Sv_2)$),
and hence $|C_3|$ is classical XOR.
We can also label $\Sv_1=\frac{1}{2}\hat x$ as 0 and $-\frac{1}{2}\hat
x$ 
as 1 for $\Sv_1$, and 
 $\Sv_2=\frac{1}{2}\hat y$ as 0 and $-\frac{1}{2}\hat y$ as 1 for 
$\Sv_2$, fixing the direction of
 $\Sv_1$ and $\Sv_2$ in $x$ and $y$ direction, respectively.
We then have $C_3(00)=C_3(11)=\frac{1}{8}$ and
$C_3(01)=C_3(10)=-\frac{1}{8}$ and this is another XOR if we read 
the sign of $C_3$ as $0$ and $1$.

%%%%%%%%%%%%%%%%

Let us see how the quantum operation works.
To remove an irrelevant degeneracy due to rotational symmetry, we fix
$\Sv_3$ in $z$-direction.
Then the quantum operator $\hat{C}_3$ reduces to 
$\hat{C}_2 \equiv \frac{1}{2}(\hat{\Sv}_1\times\hat{\Sv}_2)_z 
=\frac{i}{4}(\hat{S}_1^+ \hat{S}_2^- -\hat{S}_1^- \hat{S}_2^+)$.
The eigenvalues $\lambda$ and eigenstates (represented by 
$|S_1^zS_2^z\rangle$) of $\hat{C}_2$ are obtained as
$\lambda=0$ for $|++\rangle\equiv|0_+\rangle$ and  
$|--\rangle\equiv|0_-\rangle$, 
$\lambda=\frac{1}{4}$ for 
$\frac{1}{\sqrt{2}}(|+-\rangle+e^{-\frac{\pi}{2}i}|-+\rangle)
\equiv|R\rangle$, and 
$\lambda=-\frac{1}{4}$ for 
$\frac{1}{\sqrt{2}}(|+-\rangle+e^{\frac{\pi}{2}i}|-+\rangle)
\equiv|L\rangle$.
Note that the current states $|R\rangle$ and $|L\rangle$ correspond to
the entangled states as a result of ``square-root swap'' 
operation\cite{Makhlin99}.
As is expected from the classical picture of the current appearing when
the three spins points in $x$, $y$ and $z$ directions, it is useful to 
describe the spin state by use of different quantization axis for
$\Sv_1$ and $\Sv_2$. We choose the axis of $\Sv_1$ as in $x$-direction,
and that of $\Sv_2$ in $y$-direction.
For instance, $|0\rangle=|x\rangle$ and $|1\rangle=|-x\rangle$ for
$\Sv_1$ is written as 
$|\pm x\rangle=\frac{1}{\sqrt{2}}(|+\rangle \pm|-\rangle)$.
Then states of the two spins are expressed in terms of eigenstates of 
$\hat{C}_2$ as
\begin{eqnarray}
|\pm x,\pm y\rangle &=& \frac{1}{2}(|0_+\rangle +i|0_-\rangle) 
\pm\frac{i}{\sqrt{2}}|R\rangle \nonumber\\
|\mp x,\pm y\rangle &=& \frac{1}{2}(|0_+\rangle -i|0_-\rangle) 
\pm\frac{i}{\sqrt{2}}|L\rangle \label{map1}
\end{eqnarray}
By taking the expectation values,
we see that the classical XOR gate mentioned above is reproduced by
taking the expectation value, 
$\langle \hat{C}_2 \rangle$.

In order to implement quantum operations, we need to kill the unwanted
 state without current, $|0_\pm\rangle$.
These states carry finite total $S_z(\equiv S_z^1+S_z^2)$, 
$S_z=\pm 1$, and thus
are deleted by use of projection into $S_z=0$ subspace, 
which we write as $P_0$.
(Note that $|R\rangle$ and $|L\rangle$ are eigenstates of $S_z=0$.)
After the projection, the mapping (\ref{map1}) reduces to
\begin{eqnarray}
P_0|\pm x,\pm y\rangle &=& 
\pm\frac{i}{\sqrt{2}}|R\rangle \nonumber\\
P_0|\mp x,\pm y\rangle &=& 
\pm\frac{i}{\sqrt{2}}|L\rangle ,\label{map2}
\end{eqnarray}
and we have direct correspondence between the quantum 
spin states and two states of the current.
The operation here is a modified quantum XOR gate (neglecting the
coefficient of $\frac{i}{\sqrt{2}}$);
\begin{equation}
    \begin{array}{cccc}
  |\Sv_{1},\Sv_{2}\rangle & & C_{2} \\
    |00\rangle & \rightarrow &  |R\rangle  \\
    |01\rangle & \rightarrow &  e^{i\pi}|L\rangle \\
    |10\rangle & \rightarrow &  |R\rangle \\
    |11\rangle & \rightarrow &  e^{i\pi}|L\rangle 
      \end{array}
    \label{eq:qXOR}
\end{equation}
The extra factor of $e^{i\pi}$ can be removed by a single spin
operation if one wants.
We can easily check that this operation correctly maps
the superposition state of
the spin into the corresponding superposition state of the current.

%%%%%%%%%%%%%%%%%%%%%%%%%%%%%%%%%%%%%%%%

The operation is obviously extended to the case of more qubits.
For instance,
4-bit operation is carried out by putting five $S_i$'s on a ring, with
$S_5$ fixed in $z$-direction.
The current in this case is found 
(by a similar calculation)
to be proportional to the five-spin-chirality, $\hat C_{12345}$, 
obtained as
\begin{eqnarray}
\hat C_{12345} &=& [(\hat\Sv_1\times\hat\Sv_2)\cdot\hat\Sv_3]
(\hat\Sv_4\cdot\hat\Sv_5)
+[(\hat\Sv_3\times\hat\Sv_4)\cdot\hat\Sv_5](\hat\Sv_1\cdot\hat\Sv_2)
\nonumber\\
&&
-[(\hat\Sv_2\times\hat\Sv_4)\cdot\hat\Sv_5](\hat\Sv_1\cdot\hat\Sv_3)
+[(\hat\Sv_1\times\hat\Sv_4)\cdot\hat\Sv_5](\hat\Sv_2\cdot\hat\Sv_3).
\end{eqnarray}
We can show that this $\hat C_{12345}$ works as XOR and AND operation
combined in rather a complex way.

%%%%%%%%%%%%%%%%%%%%%%%%%%%%%%%

In the gate proposed here, the single qubit operation is achieved by
applying different magnetic field on each qubit, and for this purpose,
magnetic scanning-probe tips might be useful\cite{Loss97}.
The magnetic field to point the quantum mechanical
spin in the desired direction can be a pulse 
as in the case of pulsed NMR\cite{Gershenfeld97}.
For successive operation, one needs somehow 
to translate the quantum information carried by the current
into the spin direction, to be used as inputs of the next step
calculation, and this may be carried out by combining two rings (see below).
The present gate has a great advantage if we just want the result of a
 single operation (but on $2n$ qubits ($n\geq 1$)). 

%%%%%%%%%%%%%%%%%%%%%%%%%%%%%%%%%%%%%%%%%%%%%

\begin{figure}[btp]
\includegraphics[scale=0.5]{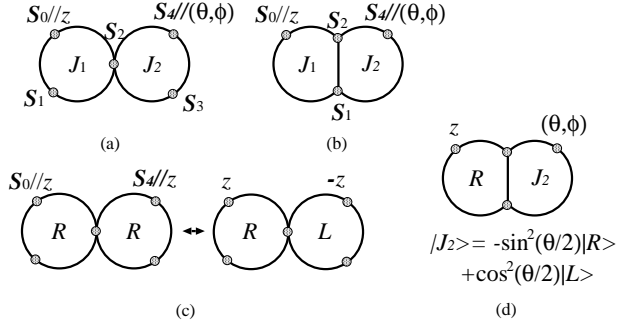}
\caption{
Two rings coupled (a) with one spin in common and (b) with two spins in
 common.
The current state $J_2$ in the second ring is a result of a
unitary transformation  of $J_1$ specified by $(\theta,\phi)$.
(c)(d): Example of operation on the flux by controling $\Sv_4$.
In (d), superposition state of current in the second ring is created
 from the $R$ state in the first ring.
\label{FIGtworings}}
\end{figure}

As is seen from the above consideration, our systems can be used as a
preparation tool of an entangled state of two or more spins.
For instance, in the case of three spins $\Sv_i$ ($i=0,1,2$), 
with $\Sv_0 // z$, we can create an entangled state of 
$|\Sv_1\Sv_2\rangle=\frac{1}{\sqrt{2}} |+-\rangle \mp i|-+\rangle$ 
by projecting the current state into $|R\rangle$ or $|\L\rangle$,
respectively. 
The current state is implemented by putting magnetic flux through the
ring (i.e. by inducing conventional persistent current)\cite{KTF03}.
By carrying out unitary transformations for the spins in the above
states, we can obtain various superposition states.  
Entangled state of three spins is also straightforward. 
We combined two rings as in Fig. \ref{FIGtworings}(a), with one spin
$\Sv_2$ in common.
Thus the current states for the first ring, $J_1$, is despribed as
$|R\rangle_1=|+-\rangle_{12} - i |-+\rangle_{12}$, and 
$|L\rangle_1=|+-\rangle_{12} + i |-+\rangle_{12}$, where
$|+-\rangle_{12}$ denotes the state of $\Sv_1$ and $\Sv_2$.
Let us point $\Sv_4$ on the second ring in arbitrary direction described
by the polar coordinates $(\theta, \phi)$.
Then the current state of the second ring is written in terms of $\Sv_2$
and $\Sv_3$ as
\begin{eqnarray}
|R\rangle_2 &=& \frac{1}{2} 
  [ \sin\theta (e^{-i\phi}|++\rangle-e^{i\phi}|--\rangle)
     -(\cos\theta+i)|+-\rangle -(\cos\theta-i)|-+\rangle]_{23} \nonumber\\
|L\rangle_2 &=& \frac{1}{2} 
  [ \sin\theta (e^{-i\phi}|++\rangle-e^{i\phi}|--\rangle)
     -(\cos\theta-i)|+-\rangle -(\cos\theta+i)|-+\rangle]_{23}
\end{eqnarray}

%%%%%%%%%%%%%%%

Thus if we prepare by use of magnetic field the 
state $|R\rangle$ for both of the rings, i.e., $|R_1 R_2\rangle$,
the realized spin state on the two rings is
\begin{equation}
|R_1 R_2\rangle=\frac{1}{2} 
[ -\sin\theta e^{-i\phi} (|+--\rangle+i|-++\rangle)
-(\cos\theta-i)|+-+\rangle +i(\cos\theta+i)|-+-\rangle
 ]_{123}
\end{equation}
and hence the entanglement of the three spins can be controled by 
$(\theta,\phi)$.
We notice that for $\theta=0$, 
$|R_1 R_2\rangle_{\theta=0}= -\frac{e^{-i\pi/4}}{\sqrt{2}} 
( |+-+\rangle + |-+-\rangle)_{123}$ and 
for $\theta=\pi$,
$|R_1 R_2\rangle_{\theta=\pi}= \frac{e^{i\pi/4}}{\sqrt{2}} 
( |+-+\rangle - |-+-\rangle)_{123}$ and this is equal to 
$-|R_1 L_2\rangle_{\theta=0}$.
{\it
This means that if we start from the state $|R_1R_2\rangle$
with $\Sv_4//z$ and flip $\Sv_4$ to be $\Sv_4 //-z$, we obtain a state
$|R_1 L_2\rangle$; the current in the second ring is reversed.
Thus the total flux created by the current is 2 in the initial state,
but is switched off to be zero by reversing $\Sv_4$; i.e., by reversing
spin we can vanish the flux even if current exist in each ring.
}(Fig. \ref{FIGtworings}(c))

Alternative way to couple two rings is to share two spins 
(Fig. \ref{FIGtworings}(b)).
In this case, the current $J_1$ and $J_2$ are both determined by $\Sv_1$
and $\Sv_2$, but the state can again controlable by $\Sv_4$.
In fact, pointing $\Sv_4//(\theta,\phi)$, the current states of the
first ring is translated into the current states of the second ring as
(after projection $P_0$)
\begin{eqnarray}
|R\rangle_1 &=& \frac{e^{i\pi/4}}{\sqrt{2}}
\left[
-\sin^2 \frac{\theta}{2}|R\rangle_2+\cos^2\frac{\theta}{2}|L\rangle_2
\right]
\nonumber\\
|L\rangle_1 &=& \frac{e^{-i\pi/4}}{\sqrt{2}}
\left[
\cos^2 \frac{\theta}{2}|R\rangle_2-\sin^2\frac{\theta}{2}|L\rangle_2
\right].
\end{eqnarray}
{\it
Thus one can create from a current in ring 1
any superposition of $|R\rangle$ and $|L\rangle$ on
the second ring.
}
(Fig. \ref{FIGtworings}(d))

%%%%%%%%%%%%%%%%%%%%%%%%%%%%%%%%%%%%%%%%%%%%%%
The readout of the target bit is carried out by measuring the flux
arising from the persistent current. 
Such measurement on a single ring 
has been successfully carried out in the case of
conventional persistent current in a ring of gold \cite{Chandrasekhar91} and 
GaAs-AlGaAs\cite{Mailly93}.
Let us give an estimate of the present effect. 
We consider as an example a ring of GaAs-AlGaAs as in
Ref.\cite{Mailly93}, where $v_F\simeq 2.6\times 10^5 $m/s, 
$\epsilon_F\simeq 1.3\times 10^{-2}$eV.
For a ring with diameter of $2\mu$m, we have 
$J\simeq 14\times(\Delta/\epsilon_F)^3$nA.
The coupling $\Delta$ depends on the distance of the conducting layer in
the semi-conductor, but for the case it is close to the interface with
the ferromagnet, $\Delta/\epsilon_F$ would be close to the value in the
ferromagnet; $\Delta/\epsilon_F\simeq 0.2$ (i.e., effective
coupling $\Delta\sim 2.6$meV).
So the current would be $0.1$nA. 
The flux due to this current is not large but
may be detected with present lock-in technique.
Much larger current would be obtained if we use 
a superconduciting ring of $p$-wave
order parameter, such as Sr$_2$RuO$_4$\cite{Maeno94}, since
the arising persistent current becomes macroscopic. 
Some semiconducting materials (like GaAs)) are known to switch to be 
ferromagnetic when magnetic impurities are doped; (Ga,Mn)As\cite{Ohno98}.
Such host materials would show a high polarizabiblity when in contact
with ferromagents, and thus  
 would be suitable for the experimental realization of the
present effect, because the coupling $\Delta$ will increase and thus the
value of the current. 

Another good way to measure the current would be to measure the Hall
like effect in the four terminal measurement.
In the presence of flux (or persistent current), the four-terminal 
conductance through a ring is expected to be asymmetric with respect to
the flux, and a finite difference of the conductance arises when 
the voltage and current leads are reversed\cite{Buttiker86}.
The difference (which may be regarded as a ``Hall conductance'', $G_H$)
is expected in our system to be 
$G_H\simeq \frac{e^2}{h}(\Delta/\epsilon_F)^3 {C}_{3}$($\sim
e^2/h\times O(10^{-2})$) for the above estimate and if $C_3\sim O(1)$. 
This is of order of typical atomic size contacts of semiconductors, and
would be measureable.
The electric measurement, being very sensitive, 
 detection of very small spin chirality $C_{3}$ would be possible, 
as well as the system with smaller coupling $\Delta$.

%%%%%%%%%%%%%%%%%%%%%%%%%%%%%%%%%%%%%%%%%%%%%%

We have demonstrated that by manupilation of spin, we can control the
persistent current in small rings.
The quantum current states are described as entangled states of two 
or more spins.
By use of coupling of two or more rings, 
unitary transformations can be carried
out on the current states and superposition states can be prepared.
Experimental demonstration of this ``quantum toy'' would be interesting,
because this can be used as an unit for quantum computing.
Implementation by use of 
rings of semiconductors or  p-wave superconductors would
be in particular interesting. 

%%%%%%%%%%%%%%%%%%%%%%%%%%%%%%%%%%%%%%%%%%%%%%
G. T. thanks H. Kohno for disucssion and
Ministry of Education, Culture, Sports, Science and
Technology, Japan and The Mitsubishi Foundation for financial support.
He thanks Consejo Superior de Investigaciones Cientificas for its
hospitality during his stay.

%%%%%%%%%%%%%%%%%%%%%%%%%%%%%%%%%%%%%%%%%%%%%%%%%%%%%%%%%%%%%%%%%%%%%

%%%%%%%%%%%%%%%%%%%%%%%

\begin{thebibliography}{99}
\bibitem{qspins}
The expression is the same as the classical case in Ref.\protect\cite{TK03}. 
This is because $\hat{\Sv}_i$'s commute each other at the lowest order
	in $\Delta$, since they are not coupled  among each other.
\end{thebibliography}
\end{document}